\documentclass[aps, twocolumn, showpacs, superscriptaddress]{revtex4-2}

\usepackage{graphicx}
\usepackage{amsmath}
\usepackage{amssymb}
\usepackage{physics}
\usepackage{hyperref}
\hypersetup{colorlinks=true,
	    final=true,
	    linkcolor=blue,
	    citecolor=blue,
	    filecolor=blue,
	    urlcolor=blue,
}

\usepackage{ulem}
\usepackage{xcolor}
\definecolor{orange}{RGB}{255,127,0}
\usepackage{makecell}

\begin{document}
\title{Electronic structure of higher-order Ruddlesden-Popper nickelates}

\author{Myung-Chul Jung}
\email{myung-chul.jung@asu.edu}
\affiliation{Department of Physics, Arizona State University, Tempe, AZ 85287, USA}
\author{Jesse Kapeghian}
\affiliation{Department of Physics, Arizona State University, Tempe, AZ 85287, USA}
\author{Chase Hanson}
\affiliation{Department of Physics, Arizona State University, Tempe, AZ 85287, USA}
\author{Bet\"ul Pamuk}
\affiliation{School of Applied and Engineering Physics, Cornell University, Ithaca, NY, 14853, USA}
\author{Antia S. Botana}
\affiliation{Department of Physics, Arizona State University, Tempe, AZ 85287, USA}


\begin{abstract}

We analyze the electronic structure of the recently synthesized higher-order nickelate Ruddlesden-Popper phases La$_{n+1}$Ni$_n$O$_{3n+1}$ ($n=4-6$)  using first-principles calculations. For all materials, our results show  large
holelike Fermi surfaces with $d_{x^2-y^2}$ character that closely resemble those of optimally hole-doped cuprates. For higher values of $n$, extra non-cuprate-like bands of $d_{z^2}$ orbital character appear. 
These aspects highlight that this Ruddlesden-Popper series can provide a means to modify the electronic ground states of nickelates by tuning their dimensionality.  With their similarities and differences to the cuprates, this new family of materials can potentially shed light on the physics of copper-based oxides.

\end{abstract}

\maketitle

\section{Introduction}

Nickelates have long been considered as cuprate analogs and have been extensively studied for their potential for high-temperature superconductivity \cite{KWLee2004_prb,Khaliullin2008_prl,Held2009_prl,Han2011_prl,Uchida2011_prl}. The realization of this promise came in 2019, with the observation of superconductivity in hole-doped NdNiO$_2$~\cite{Li2019_Nature}. This material contains infinite NiO$_2$ planes, in analogy to the CuO$_2$ planes of the cuprates, and realizes the difficult to stabilize Ni$^{1+}$ oxidation state, isoelectronic with Cu$^{2+}: d^9$~\cite{Anisimov1999_prb}. Subsequently, PrNiO$_2$ and LaNiO$_2$ were also found to be superconducting upon doping~\cite{Osada2020_nanolett, Osada2020_prm, Osada2021_advmat, Zeng2021a_arxiv}. This series of discoveries led to an immense amount of experimental ~\cite{Wang2021_nphys,Zeng2020_prl,Gu2020_ncomms,Xiang2021_cpl,Zeng2021_arxiv,Li2020_prl,Wang2020_prm,Lee2020_APLmater,
Fu2020_arXiv,Goodge2021_PNAS,Li2020_Commun,Hepting2020_nmat, Hepting2020_nmat,Chen2021_arxiv} and theoretical research  ~\cite{Jiang2020_PRL,Zhang2020_PRR,Liu2020_npjqm,
Jiang2019_PRB,Nomura2019_PRB,Werner2020_PRB,Wu2020_PRB,Botana2020_prx,
Ryee2020_prb,Choi2020_prr,Choi2020_prb,Karp2020_prx,Jesse2020_prb,Been2021_prx,Gu2020_Commun,Lechermann2020_PRB,Lechermann2020_PRX,Leonov2020_PRB,Wang2020_PRB,Sakakibara2020_PRL,Kang2021_PRL,Hu2019_PRR}.  
$R$NiO$_2$ ($R$ = rare-earth) materials are the $n=\infty$ members of a larger structural series represented by the general formula $R$$_{n+1}$Ni$_n$O$_{2n+2}$ where $n$ is the number of NiO$_2$ planes along the $c$-axis.
 Recently, the Nd-based $n=5$ material (Nd$_6$Ni$_5$O$_{12}$) has also been shown to be superconducting, unlocking the possibility to discover a whole new family of nickelate superconductors~\cite{mundy}.

\begin{figure}[tb]
    \centering
    \includegraphics[width=8.0cm]{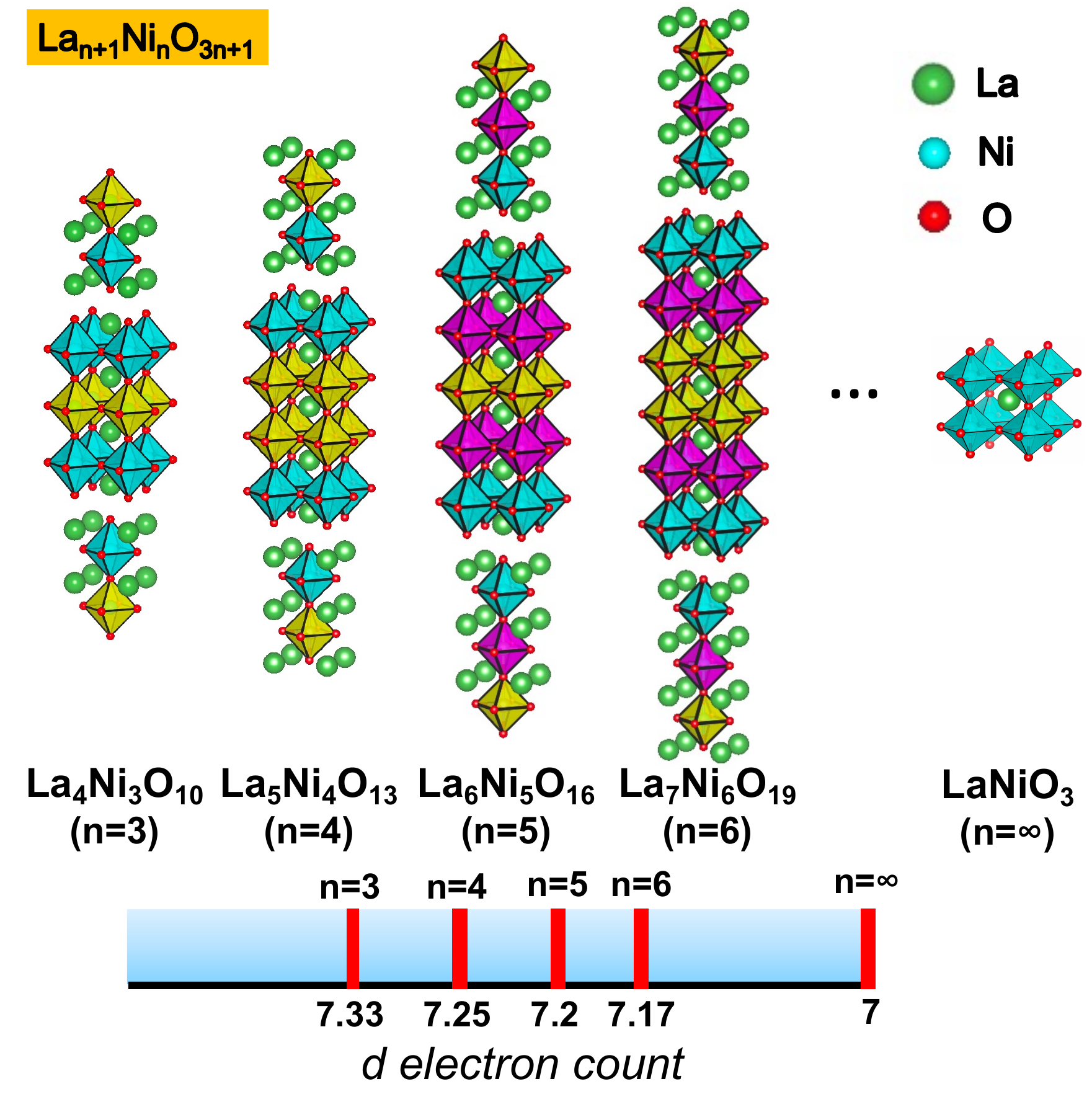}
    \caption{(Top panel) Crystal structures of the high-order RP nickelate phases ($n=3-6$) and of the perovskite nickelate ($n=\infty$). Green, cyan, and red represent the La, Ni, and O atoms, respectively. The colors of the octahedra indicate inner- (yellow), mid- (magenta), and outer-layer character (cyan)  for the corresponding Ni ions. (Bottom panel) Schematic of the Ni $3d$ electron count with increasing $n$ in these materials.}
    \label{fig1}
\end{figure}

All materials in the reduced layered $R$$_{n+1}$Ni$_n$O$_{2n+2}$ family are  synthesized by removing the apical oxygen atoms from a parent $R$$_{n+1}$Ni$_n$O$_{3n+1}$ perovskite ($n = \infty$) or Ruddlesden-Popper (RP) phase ($n$ $\neq \infty$) ~\cite{Lacorre1992_jssc,Greenblatt1997,Poltavets2007_Ln4Ni3O8,Li2020_apl_RP,Lei2017_npjqm}. These $R$$_{n+1}$Ni$_n$O$_{3n+1}$ compounds are formed by $n$-perovskite layers separated by rocksalt $R$-O spacer layers (see Fig. \ref{fig1}). 
The $R$-O layer tunes the dimensionality of
the materials as $n$ decreases: from three-dimensional (3D)-like in the $n = \infty$ case, to  quasi-two-dimensional in the $n = 1$ compound.
The parent RP phases, in particular the $n=3$ counterparts ($R$$_4$Ni$_3$O$_{10}$) at $d^{7.33}$ filling, are interesting in their own right as they present remarkable physics and notable analogies to the cuprates~\cite{Li2017_ncomms,Zhang2020_ncomms}.
Specifically, angle-resolved photoemission spectroscopy (ARPES) experiments have shown that the metallic trilayer nickelate La$_4$Ni$_3$O$_{10}$ exhibits a holelike Fermi surface of $d_{x^2-y^2}$ orbital character that closely resembles optimally hole-doped copper oxides even though, in contrast to cuprates, an extra band of $d_{z^2}$ orbital character arises~\cite{Li2017_ncomms}.
In addition, La$_4$Ni$_3$O$_{10}$ exhibits an unusual metal-to-metal transition that has been attributed to an incommensurate density wave with both charge and spin characters~\cite{Zhang2020_ncomms}.
In this context, it becomes interesting to analyze the electronic structure of higher-$n$ RP nickelates and scrutinize their similarities with and differences to the cuprates. La-based $n=4$ and $5$ RP phases (La$_5$Ni$_4$O$_{13}$ and La$_6$Ni$_5$O$_{16}$, respectively) have already been synthesized \cite{Li2020_apl_RP,Lei2017_npjqm}. These materials, with formal oxidation states between 2+ and 3+, are correlated metals and as such should offer a playground for new physics to be discovered \cite{Li2020_apl_RP}. The $n=6$ material La$_7$Ni$_6$O$_{19}$, even though not yet experimentally reported, should also be interesting to analyze trends towards the three-dimensional limit.

Here, using first-principles calculations, we study the electronic structure of La-based higher-order  RP nickelates ($n=4-6$) and perform a comparison to the known $n=3$ counterpart. 
The organization of our paper is as follows. In Sec. \ref{section2}, the structures for the higher-order RP phases ($n=4-6$) are addressed.  In Sec. \ref{section3}, we revisit the electronic structure of the $n=3$ material and present our main results on the electronic structure of the $n=4-6$ RP phases. We find that the similarities with cuprates in the fermiology of the trilayer material are kept in the $n=4-6$ nickelate RPs while their electronic structure becomes more 3D-like for increasing $n$ values, as expected.

\section{Structures and computational details}
\label{section2}

As mentioned above, in the RP nickelate structures (Fig. \ref{fig1}), a basic building block is formed by $n$ layers of corner-shared Ni-O perovskite octahedra. Each of these $n$-layer-groups is separated from
the neighboring one along the $c$ axis by a rocksalt
$R$-O unit. This gives rise to a very small
interplanar coupling between Ni-O perovskite blocks. In their corresponding reduced layered phases, obtained after oxygen reduction, the spacer rocksalt $R$-O block in the RP nickelates turns into a fluorite-like one while the Ni-O perovskite slabs turn into NiO$_2$ layers. As $n$ increases in the $R$$_{n+1}$Ni$_n$O$_{3n+1}$ series, the nominal Ni valence state changes from 2+ to 3+ (i.e., from $d^8$ to $d^7$ filling with increasing $n$). Explicitly, the RP $n=3$ compound ($R$$_4$Ni$_3$O$_{10}$) has an average Ni valence of Ni$^{2.67+}: d^{7.33}$. The $n=4$ compound ($R$$_5$Ni$_4$O$_{13}$) corresponds to Ni$^{2.75+}: d^{7.25}$, the $n=5$ material ($R$$_6$Ni$_5$O$_{16}$) to Ni$^{2.8+}: d^{7.2}$, and the $n=6$ one ($R$$_7$Ni$_6$O$_{19}$) to Ni$^{2.83+}: d^{7.17}$ (see Fig. \ref{fig1}).   

 Given that the structures of the higher-order RP nickelates have not yet been experimentally resolved,
 we construct the La-based $n= 4-6$ phases using the structure of La$_4$Ni$_3$O$_{10}$ as a reference. La$_4$Ni$_3$O$_{10}$ crystallizes in an orthorhombic ($Bmab$) or monoclinic ($P2_1/a$) 
 $\sqrt{2}$$\times$$\sqrt{2}$ supercell of the tetragonal ($I4/mmm$) parent phase ~\cite{Li2017_ncomms,Rondinelli2018_prb, Zhang2020_ncomms,Zhang2020_prm}, the latter containing two formula units. In this work, for the $n=4-6$ RP compounds, we construct the parent tetragonal cells and we neglect any possible octahedral tiltings and further distortions for simplicity. We adopt this procedure based on the good agreement achieved in La$_4$Ni$_3$O$_{10}$ ($n=3$) between the electronic structure of the tetragonal  ($I4/mmm$)  $\sqrt{2}$$\times$$\sqrt{2}$ doubled cell without octahedral tilts and the experimental orthorhombic ($Bmab$) and monoclinic ($P2_1/a$) cells with octahedral tilts (see below, Sec. \ref{section3}).

We conducted the structural optimization of the La-based $n=3-6$ RP phases with $I4/mmm$ symmetry using the Vienna $ab$ $\it{initio}$ Simulation Package (\textsc{vasp})~\cite{vasp1993, vasp1996}, having both lattice parameters and internal coordinates relaxed. We used La to avoid problems associated with the $4f$ electrons of Pr or Nd but no large changes can be anticipated from the $R$ substitution in the electronic structures, as shown for the $R$NiO$_2$ materials in a number of works~\cite{Uchida2011_prl,Botana2020_prx,Ryee2020_prb,Jesse2020_prb,Been2021_prx}. 
The generalized gradient approximation (GGA) as implemented in the Perdew-Burke-Ernzerhof (PBE) functional was used as the exchange-correlation functional~\cite{PBE1996_prl}. An energy cutoff of 650 eV and a $k$ mesh of 8$\times$8$\times$2 were adopted with a force convergence criterion of 1 meV/\AA.
The tetragonal La-based $n=3-6$ RP structures were relaxed in the nonmagnetic state. We perform calculations in the nonmagnetic state since for the known $n=3$ material Pauli paramagnetism has been reported in a wide temperature range (4-300 K)\cite{paramagnetism}  and the first-principles nonmagnetic description appropriately accounts for the existing ARPES data \cite{Li2017_ncomms}.  The derived lattice parameters for our $n=3-6$ RP phases and the bond lengths for all structures are listed in Table \ref{table1}. Good agreement is obtained between the bond lengths of our relaxations in $I4/mmm$ symmetry and the experimental structural data with $P2_{1}/a$ and $Bmab$ symmetry for the $n=3$ counterpart \cite{Zhang2020_prm, Rondinelli2018_prb}, serving as a benchmark for our structural optimizations.

\begin{table}[bt]
\caption{Lattice parameters and bond lengths of the relaxed La-based RP tetragonal structures (in \AA). $p$ and $a$ denote the planar and apical oxygens, respectively, while $i$, $m$, and $o$ denote Ni atoms in the inner, mid, and outer layers, respectively. For the $n=3$ nickelate (La$_4$Ni$_3$O$_{10}$), the bond lengths in the experimental $Bmab$ structure are 1.92 \AA (Ni($i$)-O($p$)), 1.94 \AA (Ni($i$)-O($a$)), 1.92, 1.94 \AA (Ni($o$)-O($p$)), and 2.00, 2.17 \AA (Ni($o$)-O($a$))~\cite{Zhang2020_prm}.}

\begin{center}
\begin{tabular}{ccccccccc}\hline\hline
     \multicolumn{9}{c}{higher-order RP phases}\\
     &~& $n$=3
     &~& $n$=4 
     &~& $n$=5 
     &~& $n$=6\\\cline{2-9}
     d (\AA) &~& \makecell{$a$ (\AA)=3.84 \\ $c$ (\AA)=27.96 } 
     &~& \makecell{$a $=3.84  \\ $c$=35.75 } 
     &~& \makecell{$a$=3.84  \\ $c$=43.42 } 
     &~& \makecell{$a$=3.84  \\ $c$=51.14 } \\\hline
    Ni($i$)-O($p$) &~~& 1.92  &~~& 1.92  &~~& 1.92  &~~& 1.92 \\\hline
    Ni($i$)-O($a$) &~~& 1.98  &~~& \makecell{1.94 \\ 1.97}  &~~& 1.93  &~~& 1.93 \\\hline
    Ni($m$)-O($p$) &~~&   &~~&   &~~& 1.92  &~~& 1.92 \\\hline
    Ni($m$)-O($a$) &~~&   &~~&   &~~& \makecell{1.93 \\ 1.97} &~~& \makecell{1.94 \\ 1.97} \\\hline
    Ni($o$)-O($p$) &~~& 1.92  &~~& 1.92  &~~& 1.92  &~~& 1.92\\\hline
    Ni($o$)-O($a$) &~~& \makecell{1.93 \\ 2.14}  &~~& \makecell{1.93 \\ 2.15}  &~~& \makecell{1.93 \\ 2.14}  &~~& \makecell{1.93\\ 2.14}\\\hline\hline\\
   
\end{tabular}
\end{center}
\label{table1}
\end{table}

\begin{figure}[tb]
    \centering
    \includegraphics[width=0.85\columnwidth]{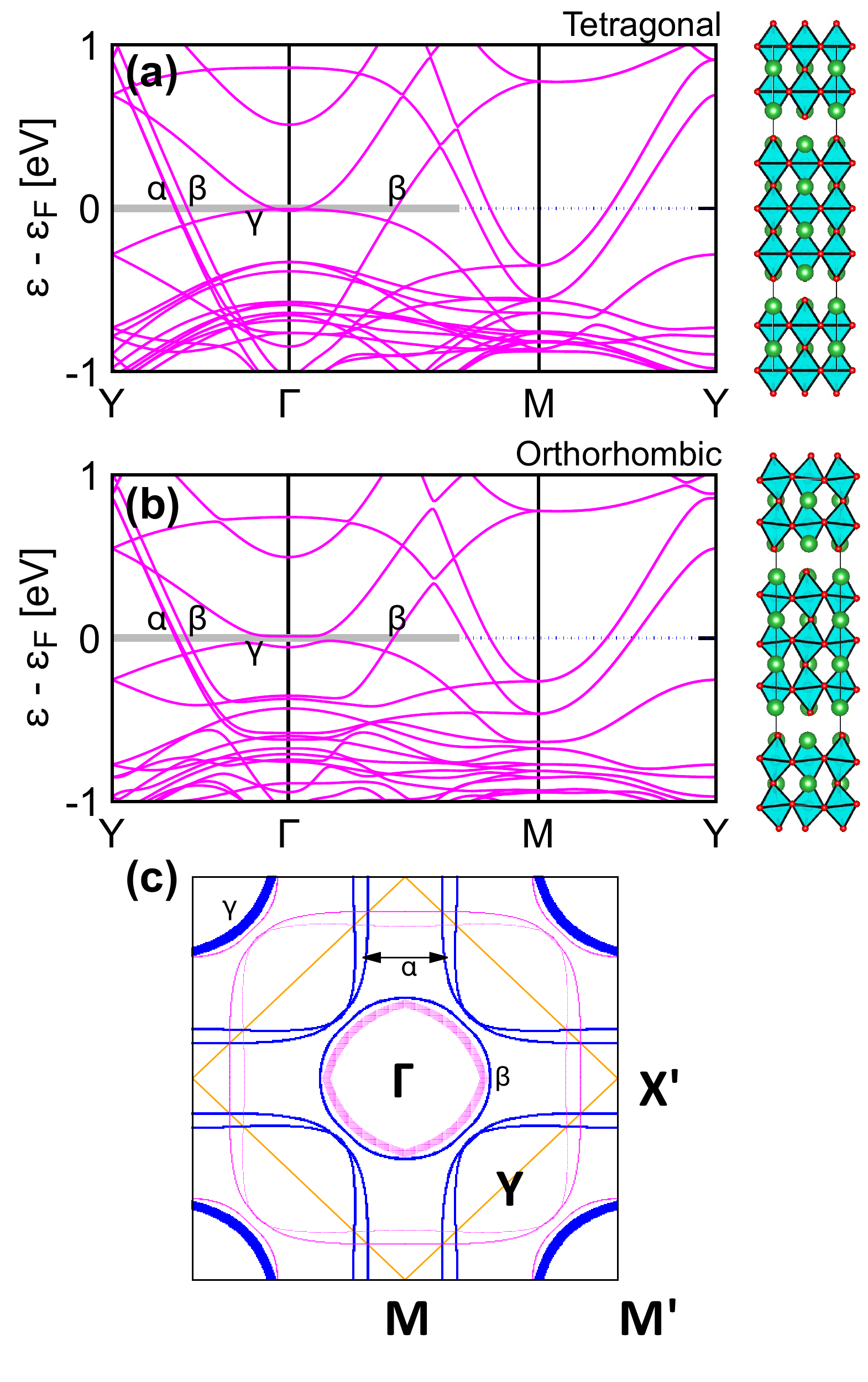}
    \caption{
    Calculated band structure of La$_4$Ni$_3$O$_{10}$ ($n=3$ RP nickelate) using (a) a doubled ($\sqrt{2}$ $\times\sqrt{2}$) tetragonal structure with $I4/mmm$ symmetry, without octahedral tilts and (b) the experimental orthorhombic structure with $Bmab$ symmetry (the band structure of the monoclinic cell with $P2_{1}/a$ symmetry is almost identical, see Appendix \ref{app_sec1}, Fig. \ref{fig5}). The bands are shown along a high-symmetry path: Y (0,1/2,0)- $\Gamma$- M (1/2,1/2,0) - Y (0,1/2,0). The labels $\alpha$, $\beta$, and $\gamma$ are used to mark the different bands contributing to the Fermi surface, following the notation of ARPES experiments reported in Ref.~\cite{Li2017_ncomms}. The gray line indicates the
blurry area from the $\gamma$ band spectral weight in the Fermi surface observed in ARPES. (c) Calculated Fermi surface of La$_4$Ni$_3$O$_{10}$. Blue color corresponds to the Fermi surface of the tetragonal (1$\times$1) $I4/mmm$ cell. Magenta color corresponds to the folded Fermi surface for the doubled ($\sqrt{2}$ $\times\sqrt{2}$) $I4/mmm$ cell.  Primed labels correspond to high-symmetry points in the 1$\times$1 tetragonal cell. }
    \label{fig2}
\end{figure}

Using the \textsc{vasp}-optimized structures, we performed further electronic structure calculations using the full-potential all-electron code \textsc{wien2k}~\cite{wien2k2020}. We continued to use  GGA-PBE as our exchange-correlation functional~\cite{PBE1996_prl}. Muffin-tin radii of 2.41, 1.94, and 1.67 a.u. were used for La, Ni, and O, respectively,  and an $R_{MT}K_{max}$=7.0 was chosen in these simulations. A fine 22$\times$22$\times$22 $k$-mesh was used for all RP phases. Additional data related to the theory is available in Ref. ~\cite{data}

\section{Electronic structure}
\label{section3}

We start by drawing a direct comparison between the known electronic structure of La$_4$Ni$_3$O$_{10}$ ($n=3$) in the experimentally resolved orthorhombic structure with $Bmab$ symmetry \cite{Li2017_ncomms,La4Ni3O10_method} and that for a $\sqrt{2}$ $\times\sqrt{2}$ supercell of our calculated tetragonal structure with $I4/mmm$ symmetry [see Figs. \ref{fig2} (a,b)].
The two band structures are very similar around the Fermi level, which serves as a further benchmark of the realistic description provided by our constructed tetragonal structures with $I4/mmm$ symmetry (see Appendix \ref{app_sec2} for further details). The different bands contributing to the Fermi surface are marked as $\alpha$, $\beta$, and $\gamma$ following the notation of Ref. \cite{Li2017_ncomms}. 
As shown in Fig. \ref{fig2}(c), La$_4$Ni$_3$O$_{10}$ displays a large hole pocket centered at the corner of the original tetragonal Brillouin zone ($\alpha$ band) of dominant $d_{x^2-y^2}$ character. In addition, there is an extra electron pocket of mixed orbital character around $\Gamma$ ($\beta$ band), and a holelike pocket of dominant $d_{z^2}$ character at the corner of the zone [$\gamma$ band in the gray area in Fig. \ref{fig2}a]. We note that the calculated Fermi surface agrees well with the ARPES data of Ref.  \cite{Li2017_ncomms}. 

\begin{figure*}
    \includegraphics[width=1.5\columnwidth]{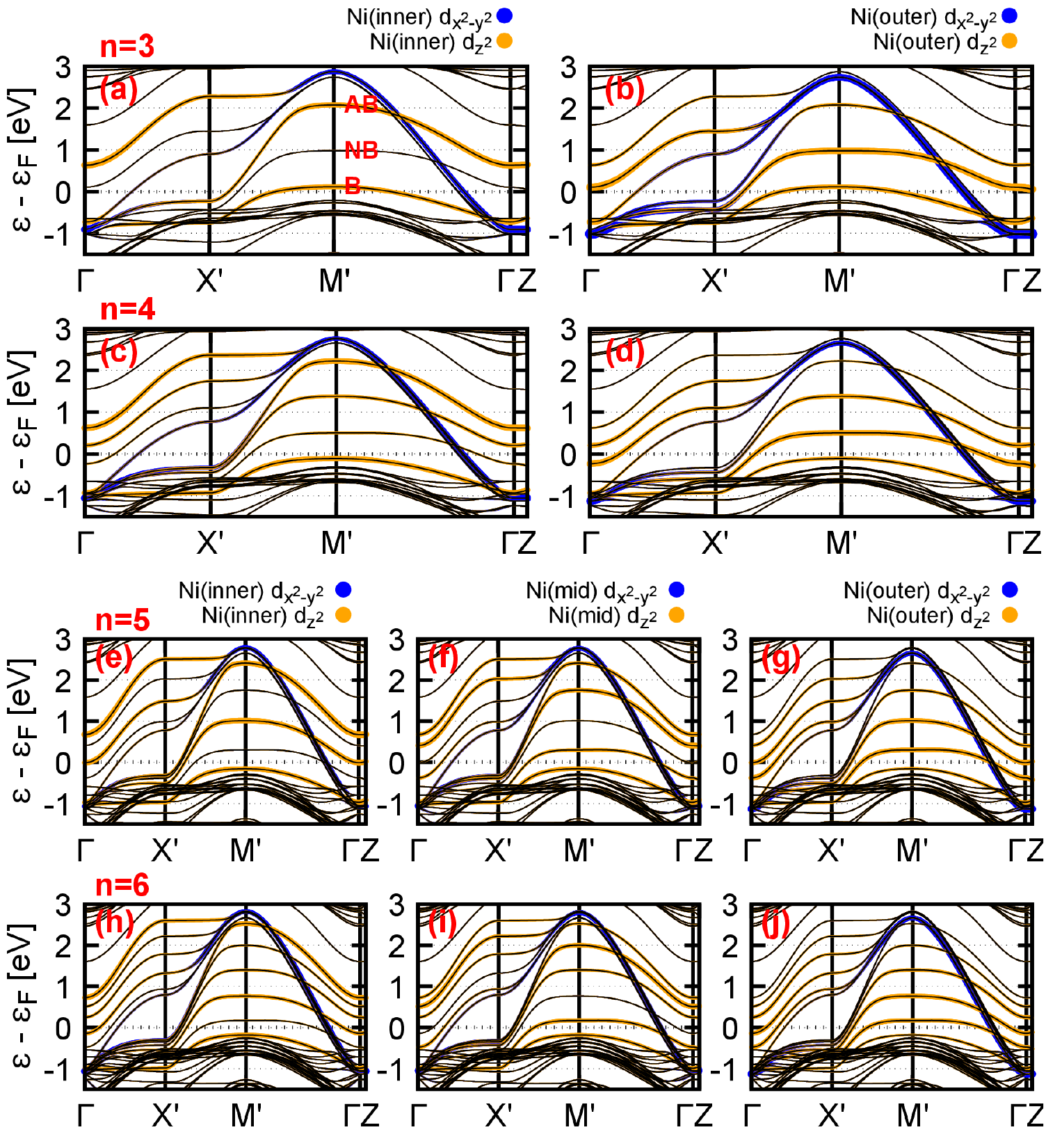}
    \caption{
  Band structures of the $n=3-6$ RP nickelates calculated in the tetragonal (1$\times$1) $I4/mmm$  cells: (a),(b) $n=3$ (La$_4$Ni$_3$O$_{10}$), (c),(d) $n=4$ (La$_5$Ni$_4$O$_{13}$), (e),(g) $n=5$ (La$_6$Ni$_5$O$_{16}$), and (h-j) $n=6$ (La$_7$Ni$_6$O$_{19}$). The bands (shown along $\Gamma$- X$^{\prime}$ (1/2,0,0)- M$^{\prime}$ (1/2,1/2,0)- $\Gamma$- Z (0,0,1/2)) are projected onto Ni-$d_{z^2}$ (orange) and Ni-$d_{x^2-y^2}$ (blue) orbitals. The (a),(c),(e), and (h) panels highlight the inner-layer Ni atoms, and the (b),(d),(g), and (j) panels the outer-layer Ni atoms for the $n=3-6$ compounds. The (f) and (i) panels highlight the mid-layer Ni atoms for the $n=5$ and $6$ compounds. The labels AB, NB, and B denote antibonding, nonbonding, and bonding Ni-$d_{z^2}$ orbitals, as described in the main text. Note that the line thickness is proportional to the contribution of each orbital.
    }
    \label{fig3}
\end{figure*}

Considering the good agreement achieved between the band structures of the calculated and experimental structures for the $n=3$ RP nickelate, we now present the electronic structure for the $n=4-6$ compounds in the tetragonal (1$\times$1) $I4/mmm$ cells. The band structures in the nonmagnetic state are shown in Fig. \ref{fig3} and they resemble that of La$_4$Ni$_3$O$_{10}$ (the corresponding densities of states are presented in Appendix \ref{app_sec3}). All the band structures ($n=4-6$) show the involvement of bands of both $d_{z^2}$ and $d_{x^2-y^2}$ character around the Fermi level, as expected from their $d^{7+\delta}$ filling. The number of $d_{z^2}$ and $d_{x^2-y^2}$ bands crossing the Fermi level increases with $n$, to account for the increasing number of Ni atoms in the structure (the band structure for LaNiO$_3$ is shown in Appendix \ref{app_sec4} as a reference for the 3D limit). 
We note that metallic behavior has been experimentally found in $n=3-5$ La-based RP nickelates, as mentioned above~\cite{Li2020_apl_RP}.
Whereas the $n=3$ system shows a kink in the $T$-dependent resistivity (a signature of density wave formation)~\cite{Sreedhar1994_jssc,Zhang1995_jssc,Li2017_ncomms,Zhang2020_ncomms}, no upturn or kink has been experimentally  found in the $n=4-5$ materials to date~\cite{Li2020_apl_RP}.

In order to clearly understand the band fillings in each material, the band structures in Fig. \ref{fig3} not only have their Ni-$d_{z^2}$  and $d_{x^2-y^2}$ orbital character highlighted, but they are also presented separately for each inequivalent Ni atom in the structure (inner, mid, and outer layers). It should first be noted that the $d_{z^2}$ bands are filled quite differently for different Ni atoms in the structure. In contrast, no relevant distinction is observed for the filling of the $d_{x^2-y^2}$ orbitals. We invoke the molecular subband picture used in Ref. \cite{Pardo2010_prl} to explain how the differences in $d_{z^2}$ filling come about. In Ref. \cite{Pardo2010_prl}, this model was applied to the electronic structure of the layered $n=3$ material, La$_4$Ni$_3$O$_8$. The $n$ molecular subbands arise as the $d_{z^2}$ orbitals hybridize strongly along the $c$ axis due to the natural quantum confinement in the structure provided by the blocking fluorite slabs between NiO$_2$ trilayers. As such, for the $n=3$ reduced compound, for each Ni triple along the $c$ axis with neighbors coupled by the hopping integral $t$, the eigenvalues and eigenvectors are  \cite{Pardo2010_prl}
\begin{equation}
\begin{aligned}
E_{z^2} &= 0,~\pm\sqrt{2}t, \\
\ket{E_{z^2}} &= \frac{1}{\sqrt{2}}(1,0,-1), ~ \frac{1}{2}(1,\mp\sqrt{2},1)
\end{aligned}
\label{eq1}
\end{equation}
The odd symmetry nonbonding (NB) state, which does not involve the inner Ni site, as shown in the eigenvector coefficients, is surrounded below and above by even symmetry bonding (B) and antibonding (AB) states. 

\begin{table}[h]
    \caption{Eigenvalues and eigenvectors of a simple molecular orbital problem with strong $d_{z^2}$ coupling for the $n=4-6$ RP nickelates. The order for the eigenvector coefficients is lowermost-outer, lowermost-mid, inner, uppermost-mid, and uppermost-outer Ni (see Fig. \ref{fig1}).}
    \begin{center}
    \begin{tabular*}{\columnwidth}{l@{\extracolsep{\fill}}ccc}
    \hline
    \hline 
        &  Eigenvalue $E_{z^2}$ & Eigenvector $\ket{E_{z^2}}$ \\
    \hline
    \textbf{$n=4$}  & &\\
     & $\frac{1}{2}(-1-\sqrt{5})t$  & $\frac{1}{\sqrt{5+\sqrt{5}}}(1,1+\sqrt{5},1+\sqrt{5},1)$ \\
       & $\frac{1}{2}(1-\sqrt{5})t$  & $\frac{1}{\sqrt{5+\sqrt{5}}}(1,-1+\sqrt{5},1-\sqrt{5},-1)$ \\
      & $\frac{1}{2}(-1+\sqrt{5})t$ & $\frac{1}{\sqrt{5+\sqrt{5}}}(1,1-\sqrt{5},1-\sqrt{5},1)$ \\
       & $\frac{1}{2}(1+\sqrt{5})t$   & $\frac{1}{\sqrt{5+\sqrt{5}}}(1,-1-\sqrt{5},1+\sqrt{5},-1)$\\
      \hline
 \textbf{$n=5$}  & &\\
  & $-\sqrt{3}t$ &  $\frac{1}{2\sqrt{3}}(1,\sqrt{3},2,\sqrt{3},1)$ \\
   & $-t$ &  $\frac{1}{2}(1,1,0,-1,-1)$ \\
    & 0 &  $\frac{1}{2}(1,0,-1, 0,1)$ \\
     & $t$ &  $\frac{1}{2}(1,-1,0,1,-1)$ \\
      & $\sqrt{3}t$ &   $\frac{1}{2\sqrt{3}}(1,-\sqrt{3},2,-\sqrt{3},1)$\\
\hline
 \textbf{$n=6$}  & &\\     
  & $-1.8t$ & $\frac{1}{4.31}(-1, 1.80, -2.25, 2.25, -1.80, 1)$\\
  & $-1.25t$ &  $\frac{1}{2.39}(1, -1.25, 0.55, 0.55, -1.25, 1)$\\
     & $-0.44t$ &  $\frac{1}{1.92}(-1, 0.44, 0.80, -0.80, -0.44, 1)$\\
    & $0.44t$ & $\frac{1}{1.92}(1, 0.44, -0.80, -0.80, 0.44, 1)$\\
     & $1.25t$ &  $\frac{1}{2.39}(-1, -1.25, -0.55, 0.55, 1.25, 1)$\\
    & $1.8t$ & $\frac{1}{4.31}(1, 1.80, 2.25, 2.25, 1.80, 1)$\\
      
     \hline 
     \hline
    \end{tabular*}
    \label{table2}
    \end{center}
\end{table}

In Fig. \ref{fig3}(a) and \ref{fig3}(b) the Ni-$d_{z^2}$ characters are emphasized for inner- and outer-layer Ni atoms, respectively, in the $n=3$ RP compound, La$_4$Ni$_3$O$_{10}$. Even though this material does have apical oxygens (unlike its La$_4$Ni$_3$O$_{8}$ layered counterpart), the molecular subband picture survives enabled by the small interplanar coupling between cells due to the blocking $R$-O perovskite slab. 
The bands display both the spectrum and the inner-outer character -- the midenergy $d_{z^2}$ band (NB) has purely outer-layer Ni atom character, while the other two bands above and below (AB and B) have both inner- and outer-layer Ni atom character. 

Based on this procedure, we calculate the eigenvalues and eigenvectors for the $n=4-6$ nickelates as shown in Table \ref{table2}. The bands once again display the calculated spectrum and inner-, mid-, and outer-layer character faithfully. For the $n=4$ (La$_5$Ni$_4$O$_{13}$) nickelate, Fig. \ref{fig3}(c) shows how the inner-layer Ni atoms have less contribution to the mid-energy $d_{z^2}$ band, in agreement with the derived coefficients for their eigenvectors. In the $n=5$ compound (La$_6$Ni$_5$O$_{16}$) the inner-layer Ni atoms do not contribute to the second and fourth $d_{z^2}$ bands [Fig. \ref{fig3}(e)], whereas the midlayer Ni atoms do not contribute to the $d_{z^2}$ midenergy band  [Fig. \ref{fig3}(f)], all in agreement with the zero coefficients in their eigenvectors.
In the $n=6$ compound (La$_7$Ni$_6$O$_{19}$), the inner-layer Ni atoms have less contribution to the second $d_{z^2}$ band [Fig. \ref{fig3}(h)], the midlayer Ni atoms to the third one [Fig. \ref{fig3}(i)] and the outer-layer ones to the highest-lying $d_{z^2}$ band [Fig. \ref{fig3}(j)], in agreement with the eigenvector coefficients as well. Based on this good agreement between the band character projected onto the inner-, mid-, and outer-layer Ni atoms and the eigenvalues and eigenvectors in our model, natural quantum confinement effects (enabled by the blocking $R$-O perovskite unit) can be anticipated in these nickelate materials.
Further, this finding stresses the importance of distinguishing the inner-, mid-, and outer-layers of Ni atoms, since some charge differentiation among them will likely take place, as happens in multilayered cuprates.

\begin{figure}[tb]
    \centering
    \vspace{0.5cm}
    \includegraphics[width=0.9\columnwidth]{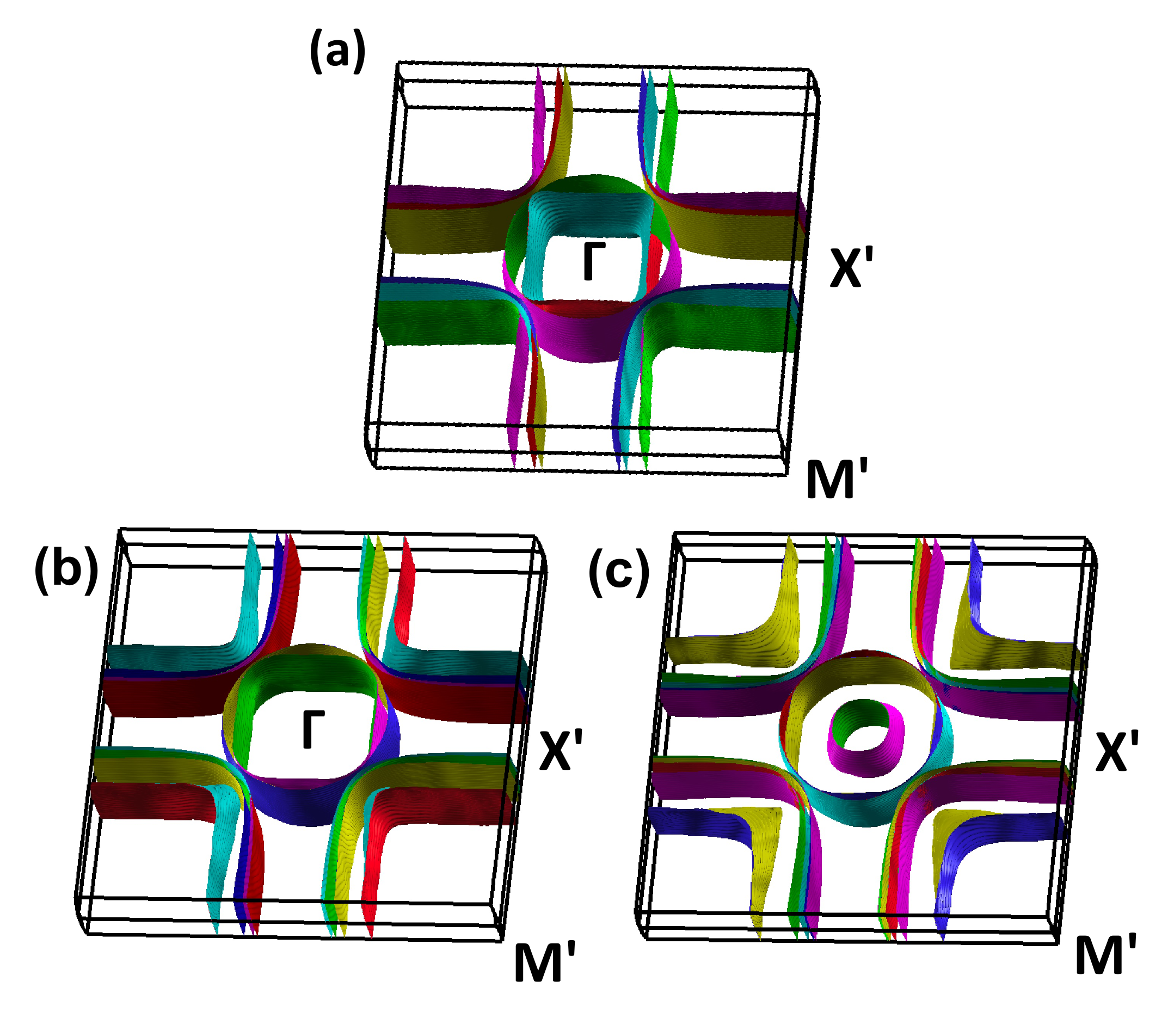}
    \caption{Fermi surfaces of the RP nickelate phases calculated in the tetragonal (1$\times$1) $I4/mmm$ structure for (a) $n=4$ (La$_5$Ni$_4$O$_{13}$), (b) $n=5$ (La$_6$Ni$_5$O$_{16}$), and (c) $n=6$ (La$_7$Ni$_6$O$_{19}$). 
    } 
    \label{fig4}
\end{figure}

Fig. \ref{fig4} shows the Fermi surface plots for the $n=4-6$ RP nickelates calculated in the tetragonal (1$\times$1) $I4/mmm$ cells. 
Similar to cuprates, all RP nickelates display 
large hole pockets of $d_{x^2-y^2}$ character (resembling cupratelike arcs), that are centered at the corner of the original Brillouin zone.
The number of these $d_{x^2-y^2}$ pockets is equal to the number of inequivalent Ni atoms and the Fermi surface sheets are split due to the interlayer hopping. In addition to the $d_{x^2-y^2}$ holelike sheets, extra electronlike pockets appear at the zone center of mixed $d_{z^2}$+$d_{x^2-y^2}$ character or pure $d_{z^2}$ character. There is more contribution from these central pockets as $n$ increases, in agreement with the increasing number of Ni atoms in the structure.  
Also, the fermiology can be seen to become more 3D-like as $n$ increases with an extra outermost arc of dominant $d_{z^2}$ character (associated with the outer-Ni atoms) appearing in the $n=4-6$ phases (see Appendix \ref{app_sec5} for more details on the Fermi surfaces).  The fact that holelike pockets resembling a
cuprate Fermi surface are found likely implies the possibility that more cuprate properties could be
achievable in this family of materials. 
Importantly, our results are consistent with ARPES data in La$_5$Ni$_4$O$_{13}$~\cite{Li2020_apl_RP} even though the electronlike band around the zone center is not visible in these experiments (the authors argue this is likely due to the strong $k_z$ dispersion of this band that cannot be probed by He-I$\alpha$ photons). In the trilayer system, Fermi surface nesting is a distinct possibility as shown in Ref. \cite{Zhang2020_ncomms}. Testing this possibility in higher-$n$ nickelates would require calculating the static susceptibilities, which is left for future work, but is definitely an interesting route to pursue.

\section{Summary}
\label{summary}
In summary, we have presented the electronic structure of the newly synthesized higher-order nickelate RP phases ($n=4-6$) using first-principles calculations. Based on the known electronic structure of La$_4$Ni$_3$O$_{10}$, we analyze the electronic structure and Fermi surfaces of our calculated RP phases to find their similarities with and differences to the cuprates. Our results show, for all materials, large holelike Fermi surfaces of $d_{x^2-y^2}$ character that closely resemble the Fermi surface of optimally hole-doped cuprates. However, extra $d_{z^2}$ orbitals appear with increasing involvement (and consequently 3D-like character) for higher values of $n$.  
These aspects indicate that nickelate physics could potentially shed light on the origin of certain aspects of cuprate physics. We expect that our study will stimulate further experiments to characterize the electronic structure of higher-order nickelate RP phases.

\section*{Acknowledgements}

We thank V. Pardo and G.A. Pan for fruitful discussions. We acknowledge NSF Grant No. DMR-2045826 and the ASU Research Computing Center for HPC resources. This research was supported in part by the National Science Foundation [Platform for the Accelerated Realization, Analysis, and Discovery of Interface Materials (PARADIM)] under Cooperative Agreement No. DMR-2039380.

\appendix

\section{Electronic structure of the $n=3$ $P2_{1}/a$ structure}
\label{app_sec1}
The $n=3$ material La$_4$Ni$_3$O$_{10}$ is known to crystallize in an orthorhombic ($Bmab$) or monoclinic ($P2_1/a$) structure that is a
 $\sqrt{2}$$\times$$\sqrt{2}$ supercell of the tetragonal ($I4/mmm$) parent phase ~\cite{Li2017_ncomms,Rondinelli2018_prb, Zhang2020_ncomms,Zhang2020_prm}. The electronic structure of the monoclinic ($P2_1/a$) structure for La$_4$Ni$_3$O$_{10}$ has also been calculated. The band structure is similar to that of the $Bmab$ structure shown in the main text (See Appendix Fig. \ref{fig5}). For this calculation we used the \textsc{wien2k}~\cite{wien2k2020} code with the computational parameters described in Section \ref{section2}, and a $k$-mesh of 25$\times$24$\times$9.

\begin{figure}[h]
\begin{center}
    \includegraphics[width=0.9\columnwidth]{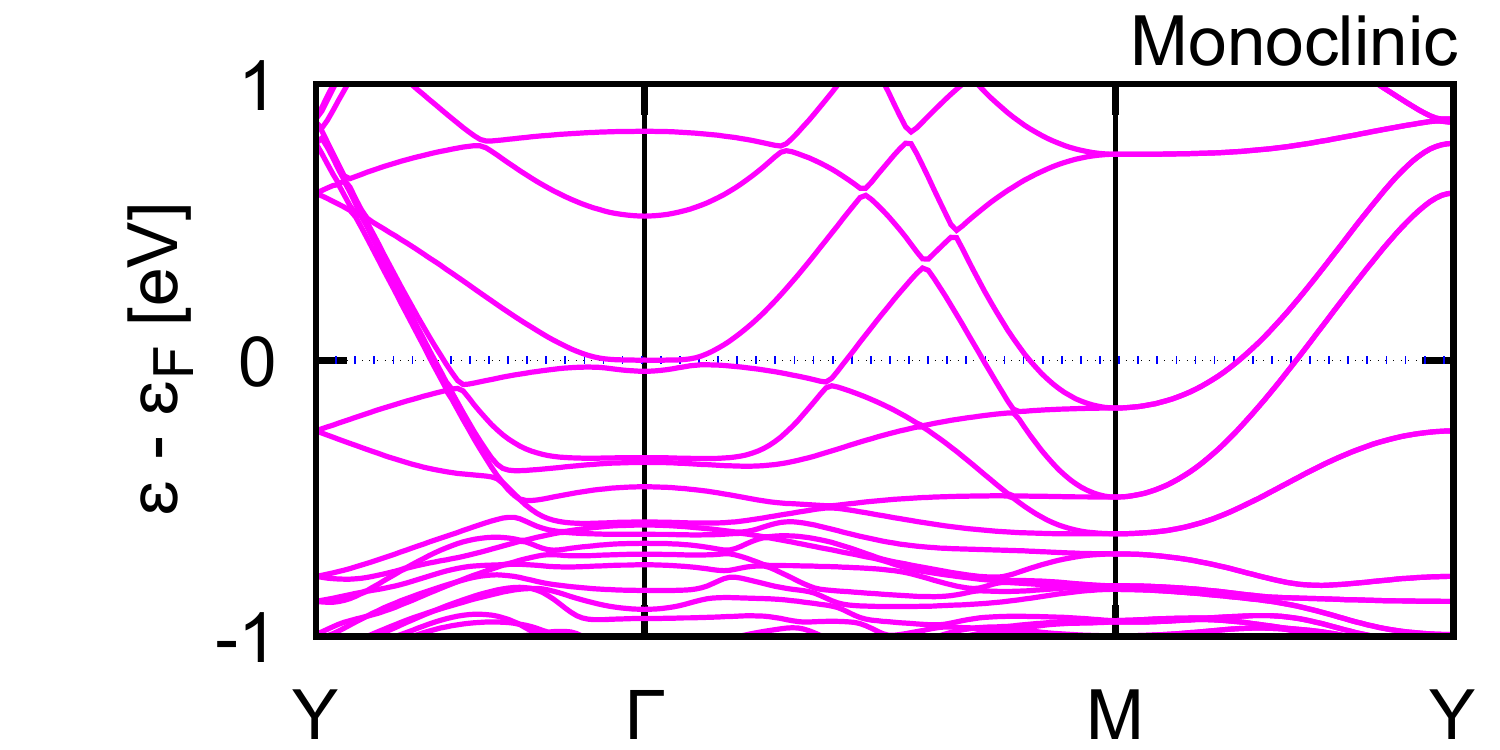}
    \caption{Electronic structure for the monoclinic structure (with $P2_{1}/a$ symmetry) of La$_4$Ni$_3$O$_{10}$. Structural data were taken from Ref. \cite{Rondinelli2018_prb}. } 
    \label{fig5}
    \end{center}
\end{figure}

\section{Electronic structure of $\sqrt{2}\times\sqrt{2}$ RP phases}
\label{app_sec2}

 In the main text, we present only the electronic structure of the calculated tetragonal (1$\times$1) $I4/mmm$ structures for the $n=4-6$ RP phases. However, we also investigated the corresponding $\sqrt{2}\times\sqrt{2}$ supercells to be able to further draw comparisons to the known electronic structure of the $n=3$ material (for which an orthorhombic ($Bmab$) and monoclinic ($P2_1/a$) symmetry have been reported, as mentioned above \cite{Rondinelli2018_prb}). We performed the corresponding electronic structure calculations in the $\sqrt{2}\times\sqrt{2}$ $I4/mmm$ cells using the \textsc{wien2k} code \cite{wien2k2020}. We employed the computational parameters described in Section \ref{section2} and used a fine $k$-mesh of 21 $\times$ 21 $\times$ 21. The obtained band structures with $d_{x^2-y^2}$ and $d_{z^2}$ characters highlighted are shown in Appendix Figs. \ref{fig6}-\ref{fig9}.

\begin{figure*}[hbt]
    \includegraphics[width=1.4\columnwidth]{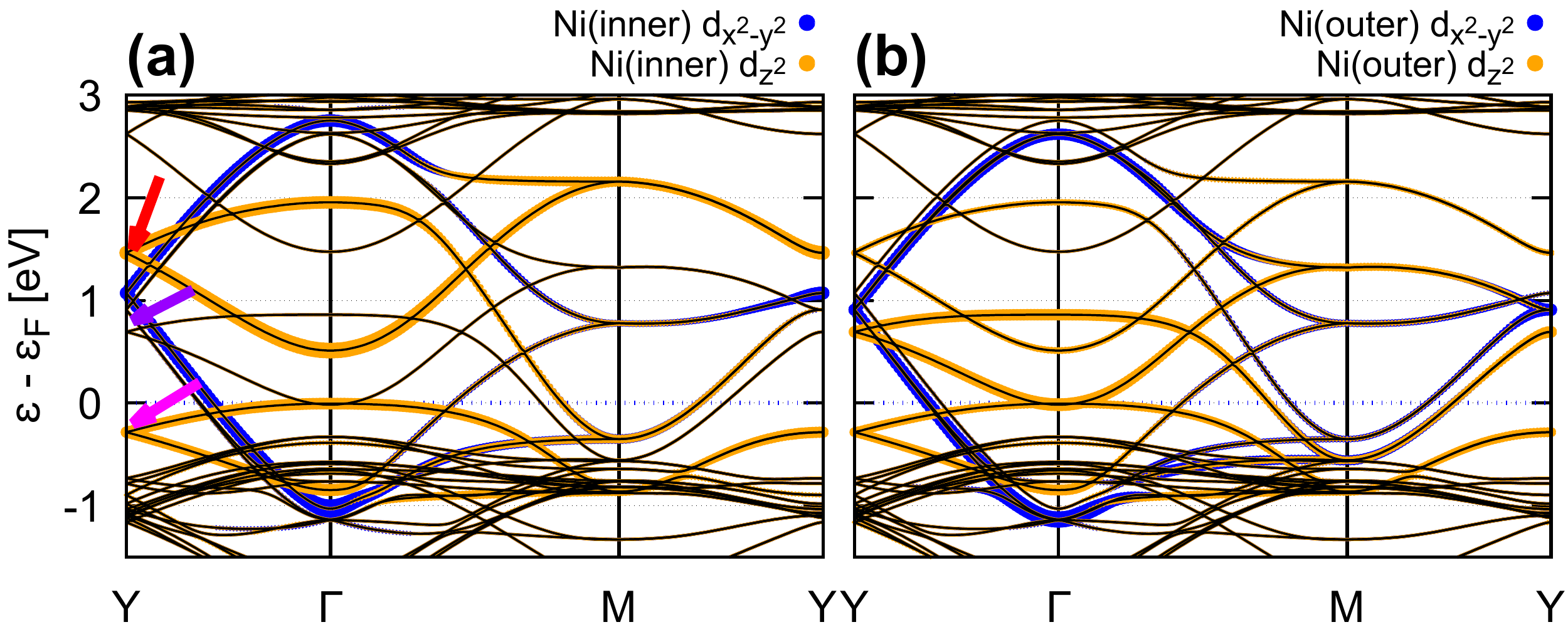}
    \caption{Orbital-resolved band structure of La$_4$Ni$_{3}$O$_{10}$ ($n=3$) in the $\sqrt{2}$ $\times$ $\sqrt{2}$ supercell of the $I4/mmm$ tetragonal structure with $d_{z^2}$ (orange) and $d_{x^2-y^2}$ (blue) characters highlighted for inner (a) and outer (b) Ni atoms.  As in Fig. \ref{fig3}(a), we denote the molecular subbands for the AB(red arrow), NB(purple), and B(magenta) $d_{z^2}$ states.} 
    \label{fig6}
\end{figure*}

\begin{figure*}[hbt]
    \includegraphics[width=1.4\columnwidth]{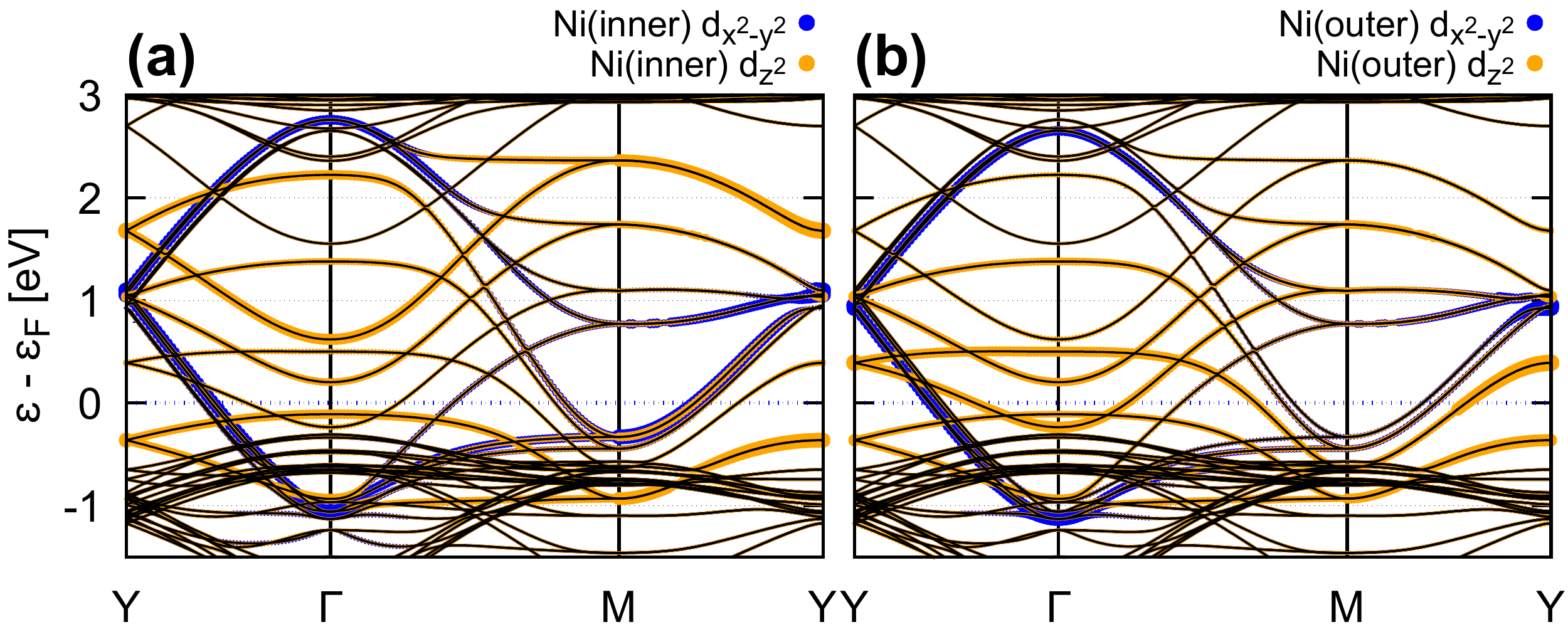}
    \caption{Orbital-resolved band structure of La$_5$Ni$_{4}$O$_{13}$ ($n=4$) in the $\sqrt{2}$ $\times$ $\sqrt{2}$ supercell of the $I4/mmm$ tetragonal structure with $d_{z^2}$ (orange) and $d_{x^2-y^2}$ (blue) characters highlighted for inner (a) and outer (b) Ni atoms.}  
    \label{fig7}
\vspace{5mm}
    \includegraphics[width=1.4\columnwidth]{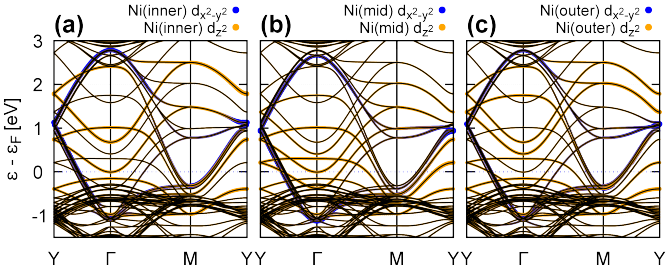}
    \caption{Orbital-resolved band structure of La$_6$Ni$_{5}$O$_{16}$ ($n=5$) in the $\sqrt{2}$ $\times$ $\sqrt{2}$ supercell of the $I4/mmm$ tetragonal structure with $d_{z^2}$ (orange) and $d_{x^2-y^2}$ (blue) characters highlighted for inner (a), mid (b), and outer (c) Ni atoms.} 
    \label{fig8}
\end{figure*}

\begin{figure*}[h]
    \includegraphics[width=1.4\columnwidth]{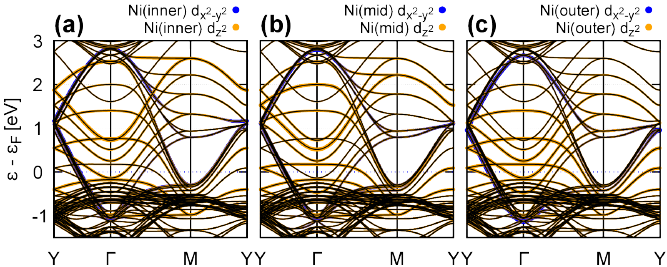}
    \caption{Orbital-resolved band structure of La$_7$Ni$_6$O$_{19}$($n=6$)  in the $\sqrt{2}$ $\times$ $\sqrt{2}$ supercell of the $I4/mmm$ tetragonal structure with $d_{z^2}$ (orange) and $d_{x^2-y^2}$ (blue) characters highlighted for inner (a), mid (b), and outer (c) Ni atoms.} 
    \label{fig9}
\end{figure*}

\section{Density of States of RP nickelate phases}
\label{app_sec3}
The atom-resolved densities of states for the tetragonal $n=3-6$ La-based RP phases are shown in Appendix Fig. \ref{fig10}. The centroids of the Ni and O states do not significantly change with $n$. 

\begin{figure*}[h]
    \includegraphics[width=1.1\columnwidth]{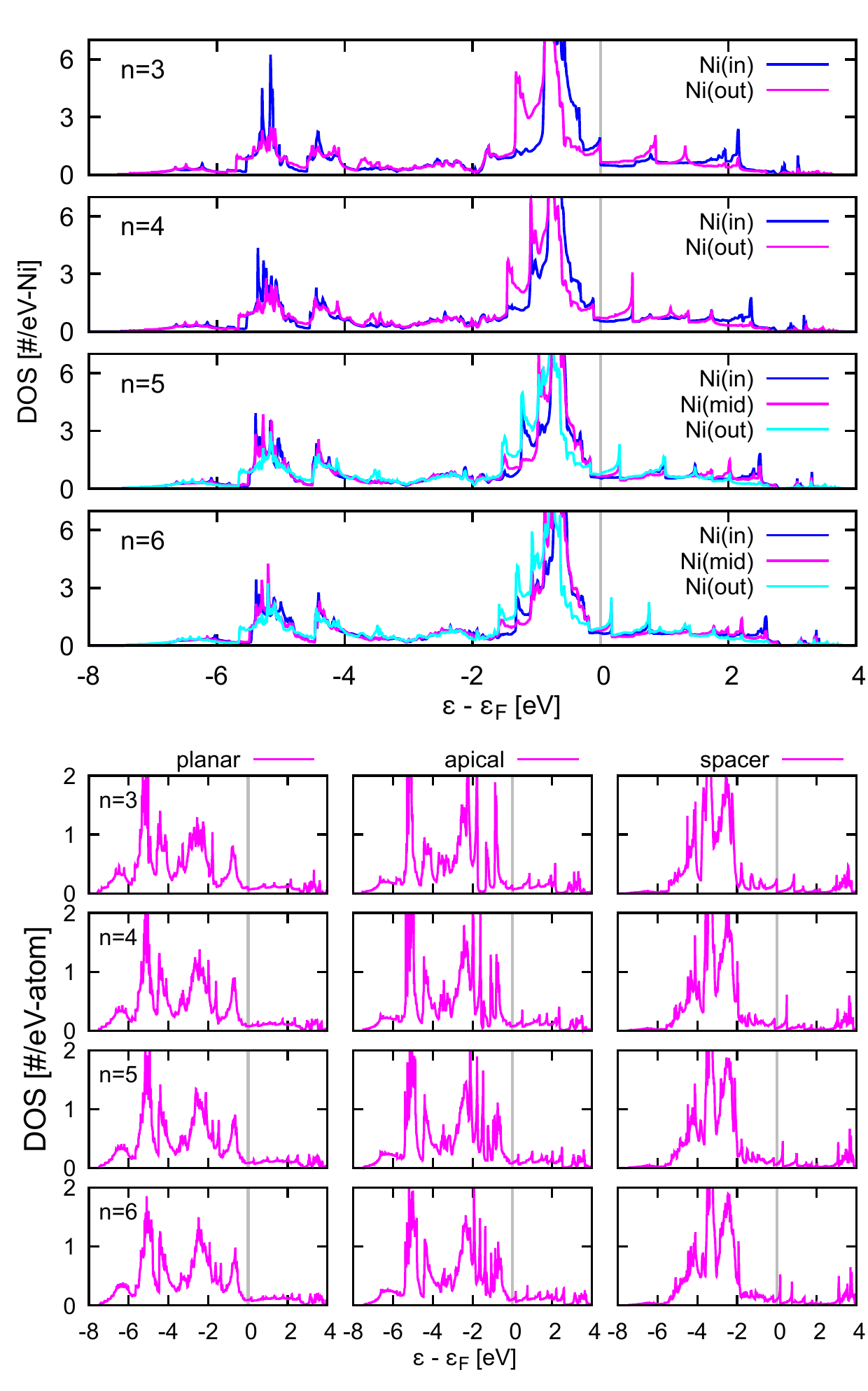}
    \caption{Atom-resolved densities of states for Ni (top) and O (bottom) ions in La$_4$Ni$_3$O$_{10}$ ($n=3$), La$_5$Ni$_4$O$_{13}$ ($n=4$), La$_6$Ni$_5$O$_{16}$ ($n=5$), and La$_7$Ni$_6$O$_{19}$ ($n=6$) RP phases obtained using the tetragonal (1$\times$1) $I4/mmm$ cells. In the density of states (DOS) plots for the oxygen atoms, spacer refers to the oxygens in the LaO blocking layer.}
     \label{fig10}
\end{figure*}

\section{Electronic structure of LaNiO$_3$}
\label{app_sec4}
The electronic structure (band structure and Fermi surface) of LaNiO$_3$ in a tetragonal $P4/mmm$ and cubic $Pm\bar{3}m$ phase obtained using GGA and the computational parameters described above are shown in Appendix Fig. \ref{fig11}. We used a $k$-mesh of 27$\times$27$\times$27 for the cubic phase  and 25$\times$25$\times$30 for the tetragonal  phase.
\\

\begin{figure*}[h]
    \includegraphics[width=1.5\columnwidth]{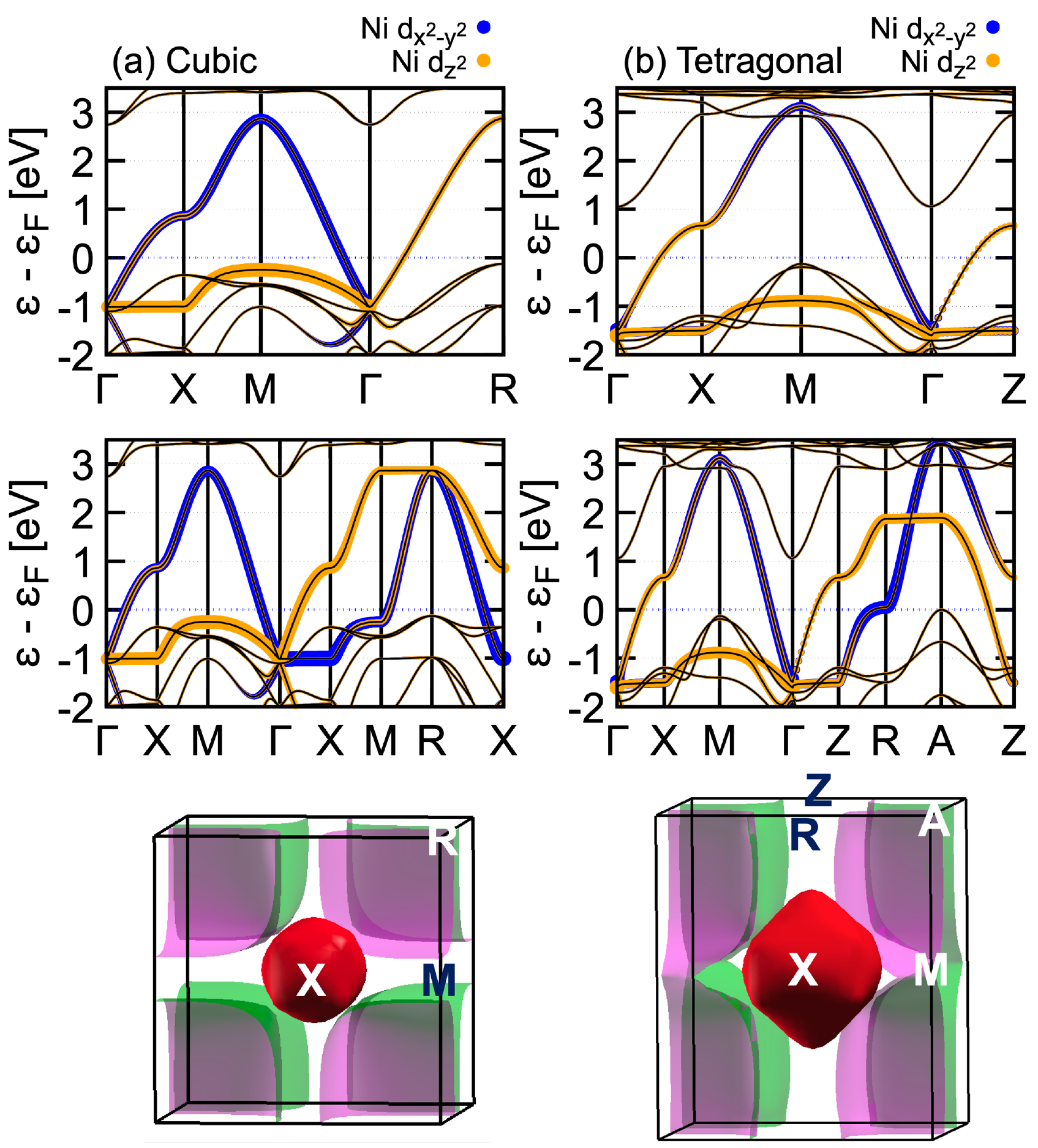}
    \caption{(Top and middle panel) Orbital-resolved band structures of LaNiO$_3$ in the cubic $Pm\bar{3}m$ (a) and the tetragonal $P4/mmm$ (b) phases, respectively, with highlighted character for Ni-$d_{z^2}$ states in orange and for Ni-$d_{x^2-y^2}$ states in blue. 
    (Bottom) Fermi surface of LaNiO$_3$ calculated in the cubic and the tetragonal cells. The employed high-symmetry paths are the following: $\Gamma$ (0,0,0), X (1/2,0,0), M (1/2,1/2,0), R (1/2,1/2,1/2) in the cubic phase and $\Gamma$ (0,0,0), X (1/2,0,0), M (1/2,1/2,0), Z (0,0,1/2), R (1/2,0,1/2), A (1/2,1/2,1/2) in the tetragonal phase.} 
    \label{fig11}
\end{figure*}

\section{Fermi surface of RP nickelates: orbital characters}
\label{app_sec5}

Here, we further analyze the character of the different Fermi surface sheets shown in Fig. \ref{fig4} of the main text for the $n=4-6$ RP nickelates. We define the corner-centered Fermi surfaces as the arcs for convenience. The $\Gamma$-centered Fermi surfaces and the corner-centered arcs are electron- and hole-like, respectively. There are Fermi arcs of dominant $d_{x^2-y^2}$ character in each material (one for  each inequivalent Ni atom in the structure).  Interestingly, extra arcs of $d_{z^2}$ character start showing up as one moves towards the 3D limit with increasing $n$, as described below.
As such, for the $n=4$ material, the outermost arc has dominant outer-Ni $d_{z^2}$ character, the middle arc outer-Ni $d_{x^2-y^2}$ character, and the inner arc inner-Ni $d_{x^2-y^2}$ character. For the electron pockets, the outer circular one has dominant inner-Ni $d_{x^2-y^2}$ character, and the inner square-like one has outer-Ni $d_{z^2}$ orbital character. The Fermi surface arcs from the corner to $\Gamma$ in the $n=5$ phase show, in sequence, outer-Ni $d_{z^2}$ character, outer-Ni, mid-Ni, and inner-Ni $d_{x^2-y^2}$ characters, respectively. The two large electron pockets centered around $\Gamma$ have inner- and mid-Ni $d_{x^2-y^2}$ character.
For the $n=6$ material, from the zone corner to $\Gamma$ the first arc shows a mixture of outer- and mid-Ni $d_{z^2}$ character, the subsequent two arcs have dominant outer-Ni $d_{x^2-y^2}$ character, followed by a mid-Ni $d_{x^2-y^2}$ arc, and finally one arc of inner-Ni $d_{x^2-y^2}$ character. The electron pockets have mid-Ni $d_{x^2-y^2}$, inner-Ni $d_{x^2-y^2}$, and inner- and outer-Ni $d_{z^2}$ characters, respectively.

\clearpage

\begin{thebibliography}{66}
\bibitem [{La4()}]{La4Ni3O10_method}%
  \BibitemOpen
  \bibinfo {note} {We investigated the electronic structure of $Bmab$ using
  $\textsc{wien2k}$. We used $k$-mesh of 21 $\times$ 21 $\times$ 21 and GGA-PBE
  exchange-correlation functional.}\BibitemShut {Stop}%
\end{thebibliography}
%

\end{document}